\documentclass[pra,twocolumn,flushbottom,amsmath,amssymb]{revtex4-1}
\usepackage{graphicx}% Include figure files
\usepackage{dcolumn}% Align table columns on decimal point
\usepackage{bm}% bold math
\usepackage{amsthm}
\usepackage{amsmath}
\usepackage{amssymb}
\usepackage{color}
\usepackage{mathtools}
\usepackage[hyperindex,breaklinks]{hyperref}

\definecolor{Red}{rgb}{1,0,0}

\def\ket#1{| #1 \rangle}

\def\ketbra#1{| #1 \rangle\langle #1 |}

\begin{document}

%\preprint{apl/123-QED}

%\preprint{}

\title{Fault-tolerance against loss for photonic FTQEC} %Title of paper
\author{Ben Fortescue}
\affiliation{Department of Physics, Southern Illinois University, Carbondale, IL 62901, USA}
\author{Sameer Nawaf}
\affiliation{Department of Physics, Southern Illinois University, Carbondale, IL 62901, USA}
\author{Mark S. Byrd}
\affiliation{Department of Physics, Southern Illinois University, Carbondale, IL 62901, USA}
\affiliation{Department of Computer Science, Southern Illinois University, Carbondale, IL 62901, USA} 

\date{\today}

\begin{abstract}
In general, fault-tolerant quantum error correction (FTQEC) procedures are designed to detect, correct, and be fault-tolerant against errors occurring within the qubit subspace.  But in some qubit implementations, additional
``leakage'' errors can occur in which the system leaves this subspace, and standard FTQEC procedures may not be fault-tolerant against such errors.  Generic methods for achieving fault-tolerance against leakage
are costly in terms of resources.

In this paper we demonstrate that for a leakage model common to many photonic gate implementations, FTQEC can be implemented with far fewer additional operations than in the generic case.

\end{abstract}


\maketitle %\maketitle must follow title, authors, abstract and \pacs

\section{Introduction}
Physical systems used for the storage, transport or processing of quantum information are vulnerable to noise introduced through interaction with the environment.
One effective means of preserving a quantum system against such noise is the use of quantum error-correcting codes (QECCs), which encode quantum information into larger physical
systems (e.g., encoding one qubit into multiple qubits) such that certain types of error can be correctly identified and corrected without damaging the encoded state in the process.
These typically have properties denoted by $[[n,k,d]]$: a code encoding $k$ qubits into $n$ with code distance $d$,
corresponding to an ability to correct an arbitrary error on $(d-1)/2$ of the $n$ qubits.

While the necessary operations for quantum error correction (QEC) can be implemented using standard physical quantum gates, any realistic implementation must take into account
that these gates themselves will be imperfect, and the QEC process can both introduce as well as correct errors.  This leads to the requirement that for effective
QEC using imperfect components, the QEC must be {\it fault-tolerant}, i.e., it should be constructed such that a single gate failure in the QEC cannot lead to errors on multiple
data qubits.  Many different QECCs with fault-tolerant constructions are known.  

QEC analysis typically considers quantum information encoded into two-level systems (qubits), where the QECC can correct some set of data errors which manifest as state
changes within the qubit subspace of the data qubits and/or any ancillary qubits used (for which it is known to be sufficient to consider
only combinations of Pauli $X$ (bit-flip) and $Z$ (phase-flip) operations).  We will refer to these errors, occurring within the subspace, as ``standard'' errors.
However, real physical implementations will, in general, also be subject to errors which take the system outside of the qubit subspace.
For example, the qubit may be represented by two energy levels of a trapped ion, or orthogonal polarizations of a photon,
but it may also undergo interactions which take it to a third energy level, or change the photon's position.  These are typically referred to as ``leakage'' errors.
A QECC which can correct, and is fault-tolerant against, qubit errors, may not have these properties with respect to leakage errors, and a gate failure within such a QECC which causes
a leakage error may lead to multiple errors (standard or leakage) on the data.

One of the most common examples of leakage is the loss of photons in photonic systems.  Photonic qubits are typically encoded via timing, polarisation or positional (dual-rail) encodings (with the latter two
freely interchangeable through polarising beam-splitters (PBSs)).  However, most optical components have some non-zero probability for loss of a photon through absorption or scattering out of the optical system.
This is not equivalent to a qubit error (e.g. absorption of a photon does not correspond to any change of polarisation), and the probability of such errors can be comparable to or even larger than those
of standard errors.

Thus, in general, QEC procedures must be modified to correct and be fault-tolerant against leakage.  A comprehensive procedure for achieving fault-tolerance in a general
leakage model was previously given by Aliferis and Terhal \cite{AT07}.  This was based on the use of ``leakage replacement units'' (LRUs), devices that would leave non-leaked qubits
unchanged, but replace any leaked qubit with a qubit in some arbitrary state, thus converting a leakage error into (at worst) a standard error on that qubit.  Various forms of LRU can be constructed;
one example would be a standard teleportation circuit.  For a given circuit, however, large numbers of LRUs may be required to achieve fault-tolerance against leakage errors, which can impose significant
additional resource costs (primarily in terms of additional gates, rather than time) to the computation.

In this paper we demonstrate that, for certain FTQEC procedures, these costs may be substantially reduced for certain general models of qubit loss.  In section \ref{sec-model} we specify
the behaviour of our model with respect to losses.  In section \ref{sec-ancilla} we show our main result of how the existing Steane, Shor and Knill ancilla techniques for distance-3 CSS codes may be minimally
modified to achieve fault-tolerance against losses which follow this model.  Finally in section \ref{sec-higher} we show how our analysis is
applicable to QEC in higher-distance codes.

\section{Loss model}\label{sec-model}
\begin{figure*}
\includegraphics[height=5cm]{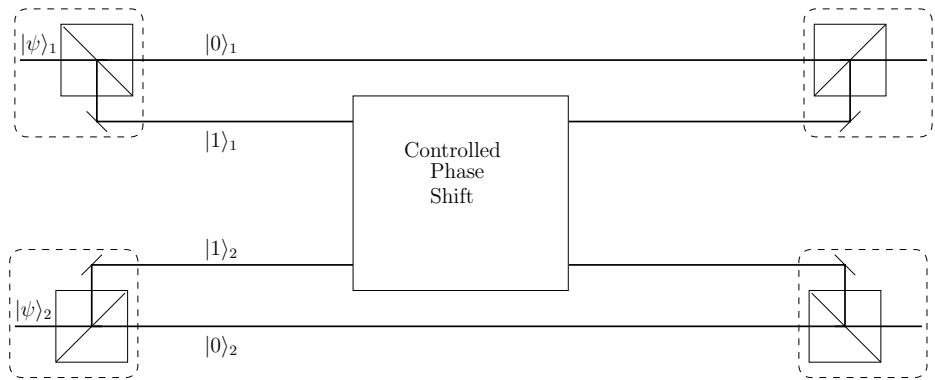}
\caption{\label{fig:2qgate} A 2-qubit phase gate satisfying our model with regard to loss errors.}
\end{figure*}

As discussed above, since loss errors fall outside of the qubit space, to understand the behaviour of a QEC circuit in the presence of loss errors requires
the description of the quantum gate behaviour to be supplemented with a description of its behaviour in the presence of loss.  We consider a model which,
in addition to being subject to standard errors, satisfies the following behaviour:

\begin{itemize}
\item Every gate (including identity/memory operations and LRUs) may undergo a fault in which the qubit(s) it acts on becomes lost (thus a two-qubit gate
can lose one or both qubits it acts on).
\item A qubit, once lost, remains lost regardless of any further operations it may undergo, with the exception of LRUs, which when correctly
operating will replace the lost qubit with one which may carry a qubit error.
\item {\bf A two-qubit operation in which one input qubit is lost will (unless an additional fault occurs)
perform the identity operation on the non-lost qubit.}   
\end{itemize}
We note that the above does not describe all behaviour with respect to lost qubits (e.g. what happens when a lost qubit is measured) but is sufficient
to demonstrate the resource reduction presented here, regardless of the unspecified behaviour.  The feature which is distinct to our model is the behaviour of the two-qubit operations.
As discussed above, one physical implementation very subject to qubit losses is the use of photonic qubits.  The above model is applicable to the large
number of proposed photonic 2-qubit gates \cite{Ralph2010209}, in which the input photonic qubits are in the dual-rail basis, and
a phase shift $\theta$ occurs if and only if both input photons are in state $\ket{1}$ and hence undergo the appropriate interaction, as depicted in Figure \ref{fig:2qgate}.  Possible means of implementing
such a phase shift include direct use of a Kerr nonlinear medium \cite{Milburn89}, mediation via an atom in cavity QED \cite{DK04}, the use of the Zeno
effect \cite{Franson04} or electromagnetically-induced transparency \cite{petro05}.  The details do not matter to our model, so long as the gate satisfies
the behaviour described above.
Such a gate, in the dual-rail basis, performs the operation
\begin{equation}
\{\ket{00},\ket{01},\ket{10},\ket{11}\}\to\{\ket{00},\ket{01},\ket{10},e^{i\theta}\ket{11}\}.
\end{equation}
Additional PBSs may be added before and after the gate to convert it to operation in the polarisation basis, as shown in the dashed boxes in Figure \ref{fig:2qgate}.
For $\theta=\pi$ this corresponds to the standard controlled-phase gate, and in the polarisation basis this can be simply converted to a CNOT gate i.e.
\begin{equation}
\{\ket{00},\ket{01},\ket{10},\ket{11}\}\to\{\ket{00},\ket{01},\ket{11},\ket{10}\}.
\end{equation}
by adding Hadamard gates in the form of half-wave plates (HWPs) applied to the target photon for the CNOT before and after the phase shift gate.

It is clear that a gate working in this manner will behave according to our model: to perform as described when both qubits are present, the state of either input qubit will
change if and only if the other qubit is present in the $\ket{1}$ channel.  Thus if either input qubit is lost, the state of the other qubit will
remain unchanged.
\subsection{Measurement}\label{sec:meas}
As mentioned above, our model does not specify the behaviour of measurement operations on lost qubits.
However, we can assume that the measurement operations we consider (projective measurements on a qubit)
have two outcomes \footnote{Additional outcomes for lost qubits, such as neither of two photodetectors in a polarisation
measurement clicking, can always be classically mapped on to one of the two outcomes if necessary.} and that a measured lost qubit will therefore
produce one of the two outcomes with some probability, independently of other qubits.

To show, with respect to fault-tolerance, that this behaviour is equivalent (at worst) to the qubit having a standard error,
we need to only show that either measurement outcome, combined with any behaviour we are considering, could occur from a qubit
with a standard error (or from no error), without any additional
errors occurring.  Note that the specific probabilities of the measurement outcomes do not have to be the same in the loss and standard error cases.

As an example (which will occur in the cases we consider), suppose we wish to show that within a procedure a loss error is equivalent to a replacement by a qubit in state $\ket{0}$
(a standard error), for a particular qubit which is then measured in the $X$ basis.  The standard error will then produce measurement outcome 0 or 1 at random (independently of other qubits)
with equal probabilities, therefore if the procedure is fault-tolerant against standard errors neither of these outcomes can lead to multiple data errors.  Hence
regardless of the specific behaviour of the measurement with respect to loss (what the probabilities of 0 or 1 are) the procedure will still be fault-tolerant in the presence of the loss error.

\section{The Steane, Shor and Knill QEC techniques}\label{sec-ancilla}
Many QECCs use one of three general ancilla techniques for QEC, which we will describe here as the Steane, Shor and Knill techniques,
and consider with regard to their fault-tolerance against loss.

\subsection{Steane QEC}
Error-correction of $Z$ errors in Steane QEC \cite{Steane97} consists of the following
\begin{itemize}
\item An encoded ancilla state $\ket{\overline{0}}$ is prepared.
\item A transversal CNOT gate is performed with the ancilla as the control and the data as the target.  This leaves the state unchanged at the logical level
but any $Z$ errors on the data are transferred to the ancilla.
\item The ancilla is then measured transversally in the $X$ basis and the error syndrome calculated.  This will be the syndrome corresponding to the $Z$ errors
on the ancilla, those acquired from the data plus any others the ancilla may have had.
\item The data is then corrected according to this error syndrome.  
\end{itemize}
Essentially the same procedure applies for correcting $X$ errors, except with preparation of a $\ket{\overline{+}}$ ancilla, which is used as the control
in a CNOT with the data as the target, and then measured in the $Z$ basis.
\begin{figure*}
\includegraphics{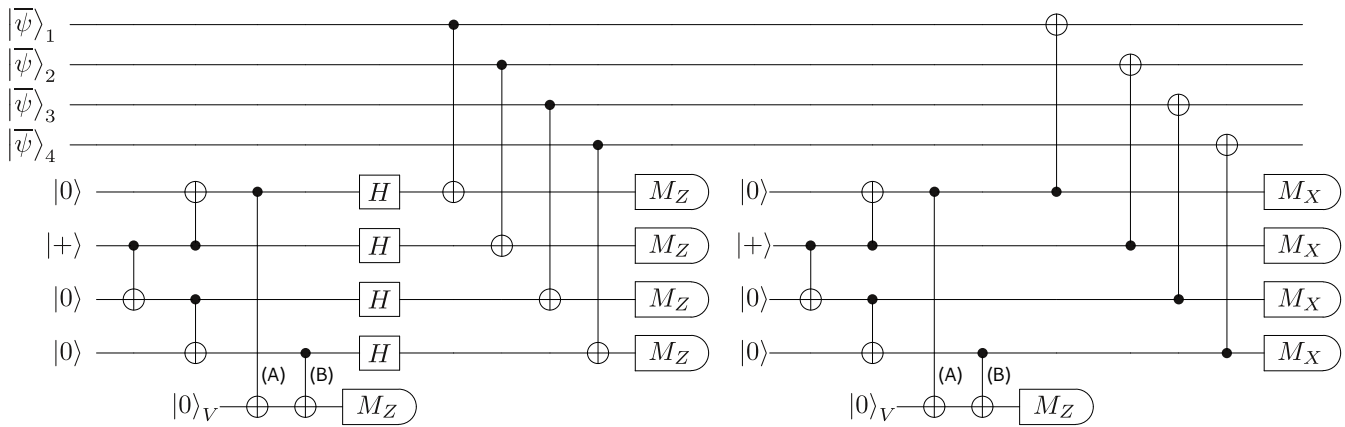}
\caption{\label{fig:shor} Measurement of two stabilisers (one a product of $X$ operators, one of $Z$) on 4 qubits using the Shor technique in a $[[7,1,3]]$ QECC.}
\end{figure*}

For a distance-3 code, fault tolerance will be violated if and only if a single gate failure leads to a logical error on the data.
The only operations performed directly on the data are the (transversal) CNOT and (if necessary) the correction operation,
which will affect 1 data qubit per fault at most.  Thus no single gate failure on these operations can lead to multiple data errors.  Similarly,
no set of errors on the ancilla which occurs after it interacts with the data can violate fault-tolerance, since these only
affect the syndrome measurement and hence the (single-qubit) correction operation.  

The only remaining possible sources
of violation are gate failures in either the ancilla preparation or the ancilla interaction with the data (i.e., the CNOT gates).
Since in the Steane QEC the ancilla measurement occurs immediately after the CNOT interaction with the data, a CNOT failure in this operation
leading to an ancilla qubit loss is equivalent to a failure leading to a standard ancilla qubit error (since both will, in general, cause measurement errors), against which the protocol is already fault-tolerant.
If the CNOT failure causes losses in both data and ancilla qubits, any incorrect syndrome due to the lost ancilla qubit 
will simply result in a correction operation on the (corresponding) lost data qubit, leading to only a single data loss error.

We are left with the case where a gate failure occurs in the ancilla preparation, leading to multiple loss and/or standard errors
on the ancilla qubits prior to interaction with the data.  This is a potential problem even in the absence of loss, hence the
use of {\it ancilla verification} in preparation.  For example, in the case of $\ket{\overline{0}}$ ancillas, multiple $X$ errors
could be transferred to the data.  Hence a second, identical ancilla is prepared (the ``verifier''), and used as the target in a transversal
CNOT from the original ancilla, then measured in the $Z$ basis.  Any $X$ errors from the original ancilla are transferred to the verifier
resulting in an error syndrome and/or logical $X$ operation on the verifier, which if detected cause the ancilla to be rejected and a new one prepared.
Ancillas must pass verification before interaction with the data (an analogous process occurs with the $\ket{\overline{+}}$), and this can be shown
to be fault-tolerant against standard errors.

More specifically, no pattern of standard qubit errors on either the newly-created
ancilla or verifier {\it alone} (regardless, in fact, of whether the pattern can be produced by a single gate failure) can result in a logical error on the data in the absence
of additional errors.   This can be seen since, in a CSS stabiliser code, any set of $X$ operations on the qubits of an encoded state can be represented as a product
of stabiliser operators, logical $\overline{X}$ operations and correctable errors.  Stabilisers can be ignored as they leave the state
unchanged, and any combination of logical operations and correctable errors on the verifier will be detected in verification in the absence
of additional errors.  Thus such a pattern on the verifier alone will result in verification failing.  Errors on the ancilla state
alone will be correctly transferred to the verifier and hence also result in failure.  (Errors on both states might not result in failure,
but we need not consider this possibility to show fault-tolerance for distance-3).  Since only $X$ errors can be transferred to the data
in the $Z$-error correction circuit, this is sufficient to show fault-tolerance (an analogous case applies to $Z$ errors in $X$-correction)
and moreover that fault-tolerance applies for any {\it arbitrary} state (without losses) of the ancilla or verifier after encoding, since in standard QEC it is sufficient
to consider $X$ and $Z$ errors alone.

\begin{figure}
\includegraphics[width=8cm]{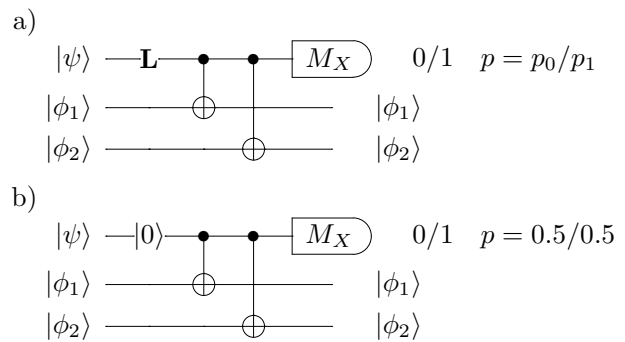}
\caption{\label{fig:loss-equiv} A qubit a) undergoing a loss error $\mathbf{L}$ can be equivalent in its behaviour to one which b) underwent the standard error of being replaced by state $\ket{0}$.
In this example the one possible difference is the values of $p_0$ and $p_1$ in the measurement of the lost qubit (undefined in our model), but as discussed in Section \ref{sec:meas} this does
not matter with respect to fault-tolerance}
\end{figure}

\begin{figure}
\includegraphics[width=8cm]{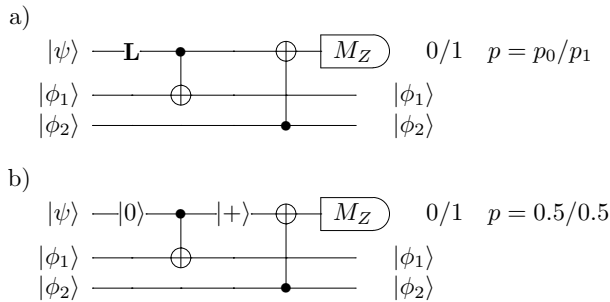}
\caption{\label{fig:loss-inequiv} If a qubit a) undergoes a loss error $\mathbf{L}$ and subsequently act as both control and target in CNOT gates, then b)
we may need to model its behaviour as multiple standard errors, in this case replacement by states $\ket{0}$ and $\ket{+}$ respectively.}
\end{figure}

We now consider the case of ancilla and/or verifier losses.  The main observation our analysis depends on is that {\bf in our loss model, the behaviour
of a lost control/target qubit in a CNOT gate (i.e., having no effect on the remaining qubit) is equivalent to its having been replaced by a qubit in state $\ket{0}$/$\ket{+}$}  Since such a replacement
is a standard single qubit error, if we can demonstrate that a qubit loss is equivalent to this replacement in all subsequent interactions,
then a standard fault-tolerant
QEC (minimally modified so that any losses on data qubits are replaced) should also be tolerant against this loss.  This is illustrated
in Figure \ref{fig:loss-equiv}.
Note that this is not always the case; for example, a qubit which
is lost and subsequently acts as both control and target in separate CNOT gates would have to be modelled as undergoing
at least two replacements, thus two separate qubit errors, which may not be a tolerable fault, as shown in Figure \ref{fig:loss-inequiv}.

Consider then a newly-created $\ket{\overline{0}}$ ancilla.  Every qubit in this ancilla undergoes three subsequent interactions: as a CNOT control
interacting with the verifier, CNOT control interacting with the data, and measurement in the $X$ basis.  Suppose some subset of these qubits have been
lost in the ancilla creation, resulting in some arbitrary state of the remaining ancilla qubits.  The lost qubits, acting as controls
in the two CNOT interactions, will behave just as though they had left the ancilla creation circuit in state $\ket{0}$.  Upon being measured in
the $X$ basis their behaviour will be no more harmful than that of a $\ket{0}$ state, which would give a result of 0 or 1 at random.
Thus the overall ancilla state is equivalent to a state with a set of standard errors after creation, against which the QEC
is already fault-tolerant.  A similar argument applies to the verifier, which acts as a CNOT target only, followed by $Z$ basis measurement, and whose losses are thus equivalent to replacement
with $\ket{+}$.  Analogous arguments apply to the $\ket{\overline{+}}$ ancillas in $X$ error correction.  Thus overall the process is fault-tolerant, and
to modify the QEC to cope with loss errors we need only add a set of LRUs on the data before the QEC.  This is contrast to the more general
model of \cite{AT07}, which also requires LRUs on all encoded ancilla states immediately after their creation.

\subsection{Shor QEC}
In the Shor QEC technique \cite{DS96}, shown in Figure \ref{fig:shor}, $n$-qubit cat states $\ket{cat_n}=\ket{0}^{\otimes n}+\ket{1}^{\otimes n}$ are used as ancillas.  Unlike in the Steane technique,
where measurement of the ancilla gives the complete error syndrome, the Shor technique provides the value of a single weight-$n$ stabiliser operator per ancilla,
a product of either all $X$ or all $Z$ operations on individual qubits (as will be the case for stabilisers in a CSS code).
\begin{figure*}
\includegraphics{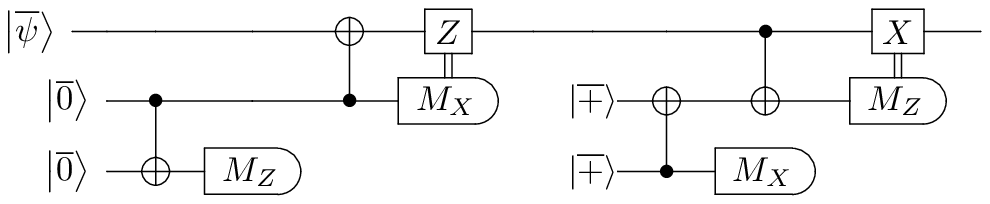}
\caption{\label{fig:steane} Fault-tolerant QEC on a data state $\ket{\overline{\psi}}$ using the Steane technique in a $[[7,1,3]]$ QECC.}
\end{figure*}
The $n$-qubit stabiliser operation $S_i$ is applied to the cat state conditioned on the state of the corresponding qubits on the data, by applying control-$X$ (i.e. CNOT)
or control-$Z$ (i.e. CPHASE gates).  Thus, for an $X$ stabiliser $S_i$ and even $n$ we have

\begin{align}
\MoveEqLeft[3] CNOT(\ket{cat_n},\ket{data})\nonumber\\
={}&\frac{1}{\sqrt{2}}\left(\ket{0}^{\otimes n}\ket{data}+\ket{1}^{\otimes n}S_i\ket{data}\right)\\
\begin{split}
={}&\frac{1}{2^\frac{n+1}{2}}\Big[(\ket{+}+\ket{-})^{\otimes n}\ket{data}\\
& +(\ket{+}-\ket{-})^{\otimes n}S_i\ket{data})\Big]\end{split}\\
\begin{split}
={}&\frac{1}{4}\Big[\ket{\textrm{even parity}}(\ket{data}+S_i\ket{data})\\
& +(\ket{\textrm{odd parity}}(\ket{data}-S_i\ket{data}\Big]\end{split}
\end{align}
where e.g. $\ket{\textrm{even parity}}$ is the equal superposition of $n$-fold tensor products of $\{\ket{+},\ket{-}\}$ containing
an even number of $\ket{+}$ states and similarly for $\ket{\textrm{odd parity}}$.  As can be seen, the states $\frac{\ket{data}\pm S\ket{data}}{\sqrt{2}}$
are the $\pm 1$ eigenstates of the $S_i$ operator, thus measurement of the parity of the ancilla state (by measuring individual
ancilla qubits in the $X$ basis) acts as a measurement of the stabiliser on the data, as required.   (Similarly, for odd $n$,
measurement of odd/even parity corresponds to $+1/-1$ eigenstates of $S_i$ respectively).

Various methods are applied to achieve fault-tolerance.  The cat states are verified before interacting with the data, to prevent single gate failures
in cat state preparation causing multiple errors which transfer to the data on interaction.  Additionally, the complete syndrome is measured three times,
and only used for error correction if at least two out of the three results agree (this prevents certain non-fault tolerant cases where the failure
of the gate interacting the data and ancilla leads to both an error on the data and an incorrect syndrome which causes a second error on correction).  
Given this feature, a single incorrect syndrome (which is the most a single gate failure can cause) cannot lead to any errors on the data.

Thus, in a model involving loss, the only remaining means by which a single gate failure can lead to multiple data errors is in the cat state
preparation and verification.  For either $X$ or $Z$
error correction, only $X$ errors from the cat-state preparation can transfer to the data, thus we consider only these errors and losses.
This leaves 5 qubit preparations and 5 CNOT gates as operations whose loss-related failures we need to investigate.  In qubit preparation operations,
we need only consider loss of the qubit.  For the CNOT gates there are 5 possible relevant failure types involving loss:
\begin{itemize}
\item Control qubit is lost.
\item Target qubit is lost.
\item Both control and target qubits are lost.
\item Control qubit is lost, target qubit has $X$ error.
\item Target qubit is lost, control qubit has $X$ error.
\end{itemize} 
We model errors occurring due to gate failure as taking place immediately after the successful operation of the gate in question  (this is without loss of generality because, for example,
an error modelled as occurring after a given gate can equivalently be modelled as occurring before the subsequent gate,
thus consideration of all possible errors in one model will also cover all possible errors in another model,
even if they are identified with different gates).  We first  consider failure of CNOT gates other than the final two (those which interact with the verifier, labelled (A) and (B) in Figure \ref{fig:shor}).

As seen from the circuit diagram, when preparing the cat state, one of the qubits is initialised in state $\ket{+}$, the others in state $\ket{0}$, and the cat state
is progressively grown from states $\ket{0}^{\otimes(m-1)}+\ket{1}^{\otimes (m-1)}$ to states $\ket{0}^{\otimes m}+\ket{1}^{\otimes m}$ by using one of the most recently-added qubits from the
$(m-1)$-qubit cat state as a source and an additional $\ket{0}$ qubit as the target in a CNOT gate.  If a qubit is lost from the cat state, the state will dephase
to an equal mixture of  $\ket{0}^{\otimes(m-1)}$ and $\ket{1}^{\otimes (m-1)}$ (which corresponds to applying an irrelevant $Z$ error with 50\% probability to one of the qubits) and any further
cat state growth involving that qubit is halted, with additional qubits remaining in state $\ket{0}$.
If a qubit is lost prior to becoming part of the cat state, no dephasing occurs, but cat state growth is similarly halted.

Thus, before interacting with the verifier, the ancilla qubit state will, in general, be a combination of a cat state of $M$ qubits (coherent or dephased), either one or two lost qubits,
and additional qubits in state $\ket{0}$.  The two qubits which interact with the verifier are the final qubits added in two independent cat state
growth processes; thus a single gate failure involving loss will cause one of them to be in state $\ket{0}$ and the other to be part of
the cat state.  Interaction of the former with the verifier does nothing, interaction of the latter causes the $\ket{0}$ state of the verifier
to be correlated with the $\ket{0}^{\otimes M}$ state of the cat state ancilla qubits.  Hence passing verification
(measuring the verifier in state $\ket{0}$) projects the ancilla qubits into a product state of $\ket{0}$ combined with the lost qubit.
These leave the data block unchanged when interacting with the ancilla, thus fault-tolerance is preserved.

If a gate failure means a control qubit receives an $X$ error and the target is lost, the construction of the circuit is such that the $X$
error will propagate no further and passing verification will again project the remaining qubits into $\ket{0}$ states.  Thus only a single
$X$ error will propagate to the data.  If the control is lost and the target has an $X$ error, any qubits the target subsequently interacts with
will be left in state $\ket{1}$.  Hence the pre-verification state will consist of a lost qubit, qubits in state $\ket{1}$ and qubits in a mixture of all $\ket{0}$
and all $\ket{1}$ (the dephased cat state from which the control was lost).  One qubit in state $\ket{1}$ and one qubit from this mixture will interact
with the verifier, hence passing verification will project all qubits in the mixture into state $\ket{1}$, leaving all ancilla 
qubits in state $\ket{1}$ apart from the lost qubit, which will consequently propagate
only a single $X$ error to the data.  Fault tolerance is again preserved.  

The remaining case is the failure of either of the final two CNOT gates, which interact with the verifier.  Gate (B) affects only a single
ancilla qubit and so cannot cause multiple data errors.  Gate (A) could in principle affect multiple ancilla qubits via the verifier,
but only if a failure affects both output qubits.  If this involves the verifier being lost, it will not affect the data further, leading to one ancilla error at most.
If the control qubit is lost but the target (the verifier) receives an $X$ error, passed verification will project the remaining qubits into a product of $\ket{1}$ states,
again leading to the one lost qubit causing a single data error.  Thus fault-tolerance is preserved in all cases.

We note that this analysis is equally applicable to $Z$ and $X$ error correction, even though the ancilla is used as the control and target respectively in the two
procedures.  When used as the target, ancilla qubits first undergo a Hadamard gate, converting any $X$ errors (against which the ancilla has been verified) to the $Z$
errors that could propagate to the data.  Similarly, with respect to data interaction, lost qubits are effectively converted between bases here as well, since
as discussed above they behave like $\ket{0}$ as a CNOT source and like $\ket{+}$ as a CNOT target.  Hence our fault-tolerance analysis for the $Z$-basis
cat state applies analogously to the $X$-basis cat state in $X$ correction.

\subsection{Knill QEC}
\begin{figure}
\includegraphics[width=8cm]{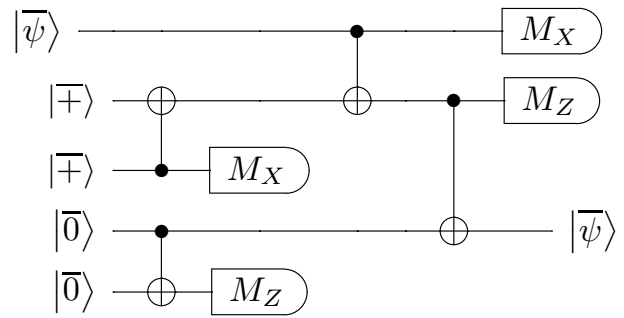}
\caption{\label{fig:knill} Fault-tolerant QEC using the Knill technique in the $[[7,1,3]]$ code}
\end{figure}

Perhaps surprisingly, the Knill QEC technique \cite{Knill05}, which seems like a natural choice in the presence of loss errors (since it replaces
the incoming data qubits with fresh physical qubits as an output), 
is an example of an otherwise-fault-tolerant QEC not remaining so in the presence of loss errors (although, as discussed below,
the additional operations required are no greater than for other techniques).  In the Knill technique, as shown in Figure \ref{fig:knill},
the QEC consists of a logical teleportation operation
on the data state, using a logical Bell state constructed from applying a CNOT from a verified (against $Z$ errors) $\ket{\overline{+}}$ state
to a verified (against $X$ errors) $\ket{ \overline{0}}$ state.  The $\ket{\overline{0}}$ state becomes the eventual data state after teleportation.

In brief, the process is fault-tolerant against standard errors because in the $[[7,1,3]]$ code, $\ket{\overline{0}}$ and $\ket{\overline{+}}$ states cannot
have multiple $Z$ and $X$ errors respectively (any incidence of multiple errors is equivalent, up to stabilizers, to a single error at most).  Thus the
only errors (in ancilla construction, all other processes being transversal) that can lead to a logical error are $X$/$Z$ errors for the $\ket{\overline{0}}$/$\ket{\overline{+}}$
states, which are checked using verification in the same manner as in the Steane technique.
However, two loss errors can occur from a single CNOT failure (if the failure causes losses on both output qubits from that CNOT) in preparation of the $\ket{\overline{0}}$ state, with the resultant
state successfully passing verification against $X$ errors.  Thus the output data state (on the same qubits) will also
have two loss errors, corresponding to a logical error in many circumstances.  We discuss in general terms (applicable to any CSS code)
how this can occur below.
\subsection{Passing verification with loss errors}
Consider a single gate failure in preparation of a $\ket{\overline{0}}_a$ state which has caused two qubit losses as discussed above, and suppose this occurs
in one of the final gates in the preparation, such that the state of the remaining qubits is
\begin{equation}
\textrm{tr}_{AB}\ketbra{\overline{0}}=\sum_{\{i,j\}\in \{0,1\}}p_{ij}\ketbra{\psi_{ij}}_a
\end{equation}
where $A$ and $B$ are the lost qubits, and
\begin{equation}
p_{ij}\ketbra{\psi_{ij}}=\langle{ij}\ketbra{\overline{0}}{ij}\rangle
\end{equation}
Consider performing (without additional errors) a verification of this ancilla against $X$ errors, by using it as the source and a
$\ket{\overline{0}}_v$ verifier state as the target in a transversal CNOT gate, after which the verifier is measured in the $Z$ basis.
In a CSS code, the target $\ket{\overline{0}}$ will be in an equal superposition of codewords of the underlying classical
code, which form a closed group under binary addition.
When applying the CNOT gate, the lost qubits will behave as though in state $\ket{0}$.  Thus the state of both ancilla and verifier
after interaction will be.
\begin{multline}
p_{00}\ketbra{\psi_{00}}_a\otimes\ketbra{\overline{0}}_v\\
+\sum_{ij\ne 00}p_{ij}\ketbra{\psi_{ij}}_a{X^i}_A{X^j}_B\ketbra{\overline{0}}_v{X^i}_A{X^j}_B,
\end{multline}
i.e., the only error-free support for the verifier state will correlate to ancilla state $\ket{\psi_{00}}$, and hence successfully passing verification will project the ancilla into state $\ket{\psi_{00}}$.  The resultant ancilla state, including the two lost qubits,
will behave, when used as a logical CNOT source, like a superposition of all codewords where qubits $A$ and $B$ are in state
$\ket{00}$, and hence will correctly leave any logical CNOT target (such as a data block) unchanged.
The behaviour is equivalent in this respect to the case
where qubits $A$ and $B$ were simply fully dephased (i.e. each underwent $Z$ errors with independent probabilities 1/2).   As in the [[7,1,3]] case, a subsequent ancilla measurement in the $X$ basis
will give an erroneous syndrome (which, however, does not violate fault-tolerance for distance-3).  An equivalent process occurs when verifying against $Z$ errors: an ancilla with lost qubits will be
projected into a superposition of codewords corresponding to the lost qubits being in state $\ket{+}$, and hence the resultant ancilla will leave a logical CNOT source unchanged when used as a target.
The consequence is that using standard verification techniques one can ``remove'' $X$ or $Z$
errors from an ancilla with lost qubits through postselection (in the sense that the postselected ancilla will not transfer such
 errors on to the data), but not both $X$ and $Z$ errors.

However, such a process will not remove the loss errors.  Thus, in the Knill technique, if the corresponding ancilla with losses passes verification, the output data state
will have multiple loss errors.  To remove this possibility, one can simply apply LRUs on each qubit of the $\ket{\overline{0}}$ state after it is created.  No further
LRUs are required as part of the QEC procedure.  Hence all three ancilla techniques can be made fault-tolerant against losses in the $[[7,1,3]]$ code by adding only 7 additional LRU operations,
compared to to 28 (Knill technique, employing post-encoding LRUs on both ancillas and verifiers) or 35 (Steane technique, as with Knill plus 7 LRUs on the data) in \cite{AT07}.

\section{Higher distance codes}\label{sec-higher}
\begin{figure*}
\includegraphics{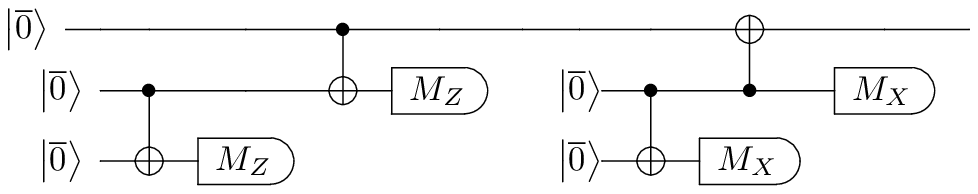}
\caption{\label{fig:golay} Verification of a $\ket{\overline{0}}$ ancilla in the $[[23,1,7]]$ Golay code}
\end{figure*}

A complete analysis of fault-tolerance in higher-distance codes requires verifying that, for a code of distance $d$ and correctable number of
errors $t=(d-1)/2$, that $l<t$ gate failures in the QEC produce no more than $l$ errors on the data block.  In general, this can be quite complex
and require multiple of rounds of verification.  We consider here the case of the Steane ancilla technique in higher-distance CSS codes, and specifically
the example of the [[23,1,7]] Golay code, for which an ancilla verification procedure (from \cite{Paetz11}) is shown in Figure \ref{fig:golay}

In general, such a technique will not be fault-tolerant against loss errors in our model.  Unlike with distance-3 codes, where errors that only
affect the syndrome measurement cannot violate fault-tolerance (since these result in at most one error on the data), a single gate failure
can result in multiple syndrome-measurement errors and consequently multiple data errors.  Similarly, since we previously only needed
to demonstrate that single failures were tolerable, it was sufficient to note that any pattern of multiple errors on an ancilla
would be detected by an error-free verification.  At higher distances one must also consider the case of errors on both the ancilla
and verifier, and the effect of different error patterns on the ancilla.

As discussed above, in the presence of loss, verification can still be used to prevent the transfer of $X$ or $Z$ errors to the data from a single ancilla, but not (without additional modification) both.
In the common higher-distance case where multiple ancilla errors
 can lead to multiple data errors due to an incorrect syndrome, the possibility of an ancilla created with multiple losses from a single gate failure would only be compatible with fault-tolerance if such ancillas were always removed in verification.
But, in general, an effective superposition of $Z$-basis codewords (i.e., a combination of lost and non-lost qubits which
behaves like such a superposition when used as a CNOT source) will have a non-zero probability to pass verification against $Z$ errors.
Thus there is a non-zero probability for an ancilla with multiple losses to pass multiple verification rounds (and lead to multiple data errors through the syndrome), even in the absence of additional gate errors, and fault-tolerance is lost in general.

We note, however, that when our loss model applies this does considerably simplify the analysis of those {\it verifier} states which,
as is common, only undergo a single interaction with the ancilla, even if the ancilla may interact with multiple verifiers as well as the data.
This is because in many ancilla creation circuits a given qubit will consistently be used as either always the source, or always the target, of a CNOT
gate.  If, for example, a verifier state is being used to test for $X$ errors, and hence interacts with the ancilla as a CNOT target, and then is
measured in the $Z$ basis, a loss error on one of these ``always target'' qubits is equivalent (or, depending on how losses are treated in measurement, no worse)
in behaviour to replacing the qubit at the point of loss with a qubit in state $\ket{+}$.  The same is true for losses of ``always source'' qubits.  Thus any such
error can be treated as a standard single-qubit error for the purpose of fault-tolerance analysis.  (We note that the $[[7,1,3]]$ Steane code is a special case of this: as discussed above, this QEC is fault-tolerant for any set of Pauli errors
on the ancilla or verifier alone, thus we only need consider the effect of loss for operations after these states are created.  For higher-distance
codes we need to consider the possibility of errors on both, hence we must consider the specific behaviour of losses within the ancilla/verifier creation circuits and the distinction between ``always source'' and ``always target'' qubits).

More generally (assuming CNOT is the only two-qubit gate) we can use our notion of $\ket{0}$/$\ket{+}$ replacement to represent any loss error occurring to qubit $q$ at point $P$ as the combination of the following:

\begin{itemize}
\item A single-qubit error occurring immediately before every CNOT gate applied to $q$ after $P$, unless immediately following a CNOT gate where $q$ acts in the same role (control or target).
\item A single-qubit error occurring before any measurement on $q$, unless that measurement is in the $X$/$Z$ basis immediately following $q$ acting as a CNOT source/target respectively.
\end{itemize}

Such a mapping will, in general, represent a single loss error as multiple non-loss single-qubit errors, and hence will demonstrate fault-tolerance against
a smaller number of loss errors than non-loss errors.  However, it provides a lower bound on the number of errors which are tolerable, higher than that
which an analysis of more general loss models would require.

\section{Conclusions}
We have demonstrated that, for a model of qubit leakage applicable to many practical two-qubit gates, especially in optical applications, standard QEC techniques
for distance-3 CSS codes can be made fault-tolerant with very few additional operations, leading to significant resource savings over
the circuits required for more general loss models.  Standard ancilla verification techniques, when passed, can effectively ``project'' lost
qubits into states which do not transfer multiple errors to the data, because the non-lost qubits are projected into a state compatible with the behaviour of the lost qubits.
While we have focused on photonic computing to motivate our loss model, we believe the same model may be applicable to trapped-ion quantum computation, which is also
subject to leakage errors \cite{knight99}, and where the precise tuning of driving pulses to atomic transitions to perform two-qubit interactions may leave qubits unchanged if interacting
with leaked qubits, as required by our model.

While higher-distance codes will generally require more additional
operations (although still fewer than a more general loss model would require), the same principles allow the analysis of fault-tolerance
to be considerably simplified compared to more general models.

\begin{acknowledgments}
The authors thank Dan Browne and Hoi Kwan Lau for useful discussions.  Supported by the Intelligence Advanced Research Projects Activity (IARPA) via Department of Interior National Business Center contract number D12PC00527. The U.S. Government is authorized to reproduce and distribute reprints for Governmental purposes notwithstanding any copyright annotation thereon. Disclaimer: The views and conclusions contained herein are those of the authors and should not be interpreted as necessarily representing the official policies or endorsements, either expressed or implied, of IARPA, DoI/NBC, or the U.S. Government.
\end{acknowledgments}

%\bibliography{aipsamp}% Produces the bibliography via BibTeX.

% \bibliographystyle{aipauth4-1}
\bibliography{loss}

\end{document}